\documentclass[preprint,aps]{revtex4}
\usepackage{graphicx}%

\begin{document}
\title{Modeling of Body Mass Index by Newton's Second Law}
\author{Enrique Canessa\footnote{E-mail: canessae@ictp.it}}
\affiliation{The Abdus Salam International Centre for Theoretical Physics,
Trieste, Italy
\vspace{3cm}
}

\begin{abstract}
Since laws of physics exist in nature, their possible relationship to
terrestial growth is introduced.
By considering the human body as a dynamic system of variable 
mass (and volume), growing under a gravity field, it is shown how 
natural laws may influence the vertical growth of humans.  
This approach makes sense because the non-linear 
percentile curves of different aspects of human physical growth
from childhood to adolescence can be described in 
relation to physics laws independently of gender and nationality.  
Analytical relations for the dependence of stature, measured 
mass (weight), growth velocity (and their mix as
the body mass index) on age are deduced with a set of common 
statistical parameters which could relate environmental, genetics 
and metabolism and different aspects of physical growth on earth.
A relationship to the monotone smoothing using functional data
analysis to estimate growth curves and its derivatives is 
established. A preliminary discussion is also presented on horizontal
growth in an essentially weightless environment ({\it i.e.}, aquatic)
with a connection to the Laird-Gompertz formula for growth.
\\
\vspace{2.0cm}
\end{abstract}
\maketitle

\section{Introduction}

The raw (and theoretically smoothed) statistical data of different 
aspects 
of human physical growth from birth to adolescence have been collected 
for decades as a function of gender and nationality.
From a practical point of view, these measurements are useful 
to monitor health care and
to highlight secular trends on the fat intake (obesity) by a 
particular population \cite{Mor00,Qhe01}.  Child growth records are also 
of special interest to Governments to enforce national health 
policies \cite{Fri90}. 

In particular, there are smoothed growth charts for boys 
and girls such as those maintained by the USA National Center 
of Health Statistics (NCHS) \cite{nchs} and the unsmoothed data 
for thousands of Japanese infants, children and adolescents 
(Hiroshima Growth Study Sample) \cite{Sumi}.  Typically, the 
charts consist of a set of non-linear percentile curves 
displaying the dependence on age of height $h(t)$, "weight" $w(t)$ 
and combinations of them such as the weight-for-height and
body mass index (BMI, {\it i.e.}, ratio 
$B(t)\equiv w/h^{2}$).  According to the
CDC growth charts \cite{nchs}, the $85^{th}$ percentile
of BMI for children is considered the overweight threshold,
and the $95^{th}$ percentile is the obesity threshold.

In constructing smoothed statistical growth curves a variety 
of asymptotic 
mathematical models have been tested to fit results for a population 
in retrospective \cite{Fle99}.  Different trial functions are
used because the data for $h$ and $w$ do not increase 
monotonically with age and because the use of 
weight-height methods (such as the BMI) affect 
directly the development of curve smoothing.
For example, modeling of human anthropometric data has been done 
in the contexts of the lambda-mu-sigma (LMS) 
model \cite{Col90}, the model functions of triple logistic 
curves \cite{Boc76}, Count-Gompertz curves \cite{Kan90,Per02} and 
Jolicoeur {\it et al.} curves \cite{Jol92}.  Another fine test 
model used is the infancy-childhood-puberty (ICP) model
\cite{Kar89}.  This model breaks down growth mathematically into
different (exponential, quadratic and logistic) functions using
different curve fitting procedures out of the growth data.
Predictions of changes in height have also been carried
out according to an empirical Bayesian approach \cite{Sho91}.
The selection of a definitive parametric approach 
for the inclusion or exclusion of some empirical data points
and to set a criteria of estimation is a topic of active
discussions because of its relevance to growth, development
and aging in living organisms \cite{Fle99}.

For open systems like those in which there are influx of mass, one 
can apply the conservation of linear momentum and energy methods 
of general physics to study their growth dynamics.  Examples in 
which momentum is gained from, or lost to, the surroundings 
include rockets \cite{Hal78,Sou02} and the falling of a 
snow ball \cite{Dia03}, respectively.  Since the human 
body is also a system of fluctuating mass (and volume),
under the influences of the acceleration of gravity $g$ and
food consumption, it is not unreasonable to consider it as 
a dynamic physics system.  One may then in principle make 
an attempt to derive a relationship based on physics laws
and observed "physical" growth. 

Since laws of physics exist in nature, their possible relationship to
human physical growth is introduced in this work.
It is shown how physics seems to influence the physical growth 
of humans.  Using Newton's second law, analytical relations 
are deduced for the time dependence of $h$, $w$, the growth velocity 
$v = dh/dt$ which depend on common parameters that could 
relate other aspects of this phenomenon such as environmental 
conditions and genetics and natural processes like metabolism, 
energy supplied by food, {\it etc}.  Within a simple physics-based 
framework, estimates of all observed human growth statistical
data including BMI on age are easily performed and are shown to fit
the observed data which makes the present model reasonable.  
A preliminary discussion on horizontal growth in an essentially 
weightless environment is also presented 
with a connection to the Laird-Gompertz formula for aquatic growth.

\section{Simple Physics Characterization}

Let us consider the human body as a system of variable mass 
$m(t)$, where mass is the amount of matter present in the
body measured using balance.  In Fig.1 the approximated center of 
mass of an upright body as seen from a particular reference frame 
at different ages is illustrated.  In scientific 
usage mass is an intrinsic property of matter and weight is 
a force that results from the action of gravity on matter.  
In everyday usage, however, weight and mass are used 
interchangeably.  The reported "weight" in the smoothed growth 
charts for humans in \cite{nchs,Sumi} corresponds to 
measured matter or mass in kilograms and not to the force of 
weight "$w(t) = m(t)g$" measured in Newtons.  In the following 
such "weight" data is referred as measured mass for the 
sake of correctness.  Changes of mass and momentum will be assumed 
to be continuous during the growing process through years from 
childhood to maturity.  

To derive the equation of motion of the body system whose mass
is not constant, let us use Newton's second law to define the 
external force $F$ on the human body. As shown explicitly 
in Appendix A, this is expressed as (see, {\it e.g.}, \cite{Hal78}) 
\begin{equation}\label{eq:force}
m\; \frac{d{\bf v}}{dt} = {\bf F}_{\it ext}  +  
    {\bf v}_{\it rel} \; \frac{dm}{dt} \;\;\; ,
\end{equation}
where $dm/dt > 0$ is the rate at which mass is gain 
({\it i.e.}, equivalent to particles entering an open system) 
and ${\bf v}_{\it rel}(t)$ is the relative velocity of the 
gained mass with respect to the body moving with a growth velocity 
of magnitude $v(t) \equiv |{\bf v}(t)|$.  The above term 
$\phi (t) \equiv {\bf v}_{\it rel}\;(dm/dt)$ corresponds to 
the momentum flux being tranferred into the body by the added
mass.  This is interpreted as the force exerted on the 
body by the mass that joins it \cite{Sou02} and ${\bf F}_{ext}$
is the external force of gravity acting on the variable mass system.  
The surface of the earth is taken as the zero level of gravitational 
potential energy.

In applying the Newton law for the vertical growth of 
humans, it corresponds to the analysis of an extended system 
of $N$ particles rather than to the analysis of a single particle
of mass $m_{i}$ and velocity ${\bf v}_{i}$ at time $t$.  
In this case one can only consider averaged values for all the
physical quantities -such as the external force, due to the complexity
of the body system.  Hence, if $n(h_{i})$ is the number 
of objects at various heights $h_{i}$, of a total of $n$ objects 
distributed among $\ell$ heigths, then the mean average value of 
any given function $f$ for those objects is in general
\begin{equation}\label{eq:average}
         \hat{f}\equiv  <f_{i}> =  
           \frac{1}{n} \sum_{i=1}^{\ell} n(h_{i}) \; f_{i} =  
           \sum_{i=1}^{\ell} \frac{n(h_{i})}{n} \; f_{i}  =
           \sum_{i=1}^{\ell} \rho_{i} \;f_{i} \;\;\;  ,
\end{equation}
where $\rho_{i}$ is interpreted as the probability that a fraction
of the objects $n(h_{i})/n$ has the height $h_{i}$. 
So as the sum of all fractions of the whole system is unity
($f=1$), the probability distribution function is also normalized 
such that $\sum_{i=1}^{\ell} \rho_{i} = 1$.
It can be seen that $\rho_{i}$ directly affect the outcome of 
averaged results since it weights the values of the given function $f$
at each height $i$.  The above mean value equation is used throughout 
the theory of statistical physics.  

Let us consider next the number of $n$ objects distributed among 
$\ell$ heights as being $N$ sets or groups of particles of added mass 
$m_{i}$ each at time $t$. Then the mean value for the external 
force of Eq.(\ref{eq:force}) in one longitudinal direction becomes
\begin{equation}\label{eq:force3}
  \hat{F}_{\it ext} \equiv <F_{i,\it ext}> =  
       \sum_{i}^{N} \rho_{i} \left\{ m_{i}\; \frac{dv_{i}}{dt} - 
              v_{i,\it rel} \; \frac{dm_{i}}{dt} \right\} \;\;\; .
\end{equation}
Statistically stationary states in which all the probabilities $\rho$ 
are time independent $\forall i$ will be only considered. 

To an observer at rest on the earth, the accumulating particles 
in the body at a given time appear to move at a velocity, say
$v_{a}(t)$, proportional to the growth velocity $v(t)$ of the 
center of mass.  For simplicity let us consider an homogenous
system so that such a proportionality is set equal to one, 
{\it i.e.} $v_{a}(t) \approx v(t)$, which corresponds to the 
maximum velocity that the added mass alone can achieve during the 
grow.  This is so because the human growth in stature is longitudinal 
until reaching a maximum height $h(\infty)$ with respect to 
earth's surface (and a maximum mass $m(\infty)$ under healthy
conditions) at adolescence (see Fig.1).  On the other hand,
to an observer on the center of mass of the body, the added 
mass appears to have a vertical motion with a velocity of magnitude 
$v(t)+v_{a}(t) \approx 2v(t)$ in the opposite direction of 
the growth with respect to the center of mass displacements. 
Therefore, for the present purposes, the relative velocity 
${\bf v}_{\it rel}(t)$ of the gained mass appears to move at 
twice the velocity of growth as seen from within the body system, 
namely $|{\bf v}_{\it rel}| = 2|{\bf v}|$.  
Hence Eq.(\ref{eq:force3}) for the mean value of the net
external force on the body as a whole is the superposition of 
contributions 
\begin{equation}\label{eq:force1}
\hat{F}_{\it ext}(t) =  <F_{i,ext}(t)> = \sum_{i=1}^{N} \rho_{i} \; 
   m_{i}(t)\; \frac{dv_{i}(t)}{dt} - 
           2 \sum_{i=1}^{N} \rho_{i} \;
     v_{i}(t) \; \frac{dm_{i}(t)}{dt} \;\;\; .
\end{equation}
with  $\rho$ the normalized probability 
distribution function of $N$ objects or set of particles 
of added mass $m_{i}$ each at time $t$.

The force of gravity $F_{g}$ due to the earth mass acting 
thoughout the human body must be also taken into account.  Similarly to 
suspended chains systems \cite{Sou02}, this force is not only 
proportional to the total weight of the body at an given height $h(t)$, 
but also to the total force due to the amount of mass at rest lying 
on the surface of the earth.  This extra term can also be understood 
by the fact that the volume (besides the mass) of the human body also
varies.  It accounts for an additional momentum that the system 
acquires as a result of changing volume (in one-dimension vertical) 
with years --see Appendix A.
In this case, the force of gravity acting on the $N$ objects of 
added mass $m_{i}$ is then given by
\begin{equation}\label{eq:force2}
\hat{F}_{g}(t) = 
\sum_{i=1}^{N} \left\{ \rho_{i} \;\frac{h_{i}(t)}{h_{i}(\infty)} m_{i}(t) \right\} g
              - w(\infty) =
  <\lambda_{i}(t)h_{i}(t)>g - m(\infty)g \;\;\; ,
\end{equation}
where $\lambda_{i}(t) \equiv m_{i}(t)/h_{i}(\infty)$ represents
mass per unit height.  The weight at rest term 
$w(\infty) = m(\infty)g = <m_{i}(\infty)>g = \sum_{i=1}^{N} \rho_{i} m_{i}(\infty)g$ 
is deduced according to the boundary condition 
$\hat{F}_{g} \rightarrow 0$ as $t \rightarrow \infty$.

\section{General Biometric formulae}

In order to derive analytical relations for the dependence of 
stature, measured mass (weight), growth velocity (and their mix 
as the body mass index) on age, let us proceed by considering 
Newton's law of motion in relation to the two distinct mean 
forces acting on the system, {\it i.e.} the net external force of 
Eq.(\ref{eq:force1}) and the gravity force of Eq.(\ref{eq:force2}).
This means to set $\hat{F}_{g} = \hat{F}_{\it ext}$, or more explicitly
\begin{equation}\label{eq:dynamics}
 \sum_{i=1}^{N} \rho_{i} \left\{ 
    m_{i}(t)\; \frac{dh^{2}_{i}(t)}{dt^{2}} -
  2 \left(\frac{dh_{i}(t)}{dt}\right) \left(\frac{dm_{i}(t)}{dt} \right)
    - \frac{h_{i}(t)}{h_{i}(\infty)} \; m_{i}(t)g \right\} 
    =  - m(\infty)\; g \;\;\; ,
\end{equation}
with $v_{i}(t) = dh_{i}(t)/dt$.
After some little algebra, and using the well-known trigonometric 
relation: $sech^{2}+tanh^{2}=1$, it is straightforward to check that 
this second order differential equation has the general solutions
\begin{equation}\label{eq:height}
h(t) =  <h_{i}(t)>  = \sum_{i=1}^{N} \rho_{i} h_{i}(\infty)\; 
     tanh \left(\frac{t-t_{i}}{\tau_{i}}\right) \;\;\; ,
\end{equation}
such that 
\begin{equation}\label{eq:height1}
h_{i}(\infty) = \frac{1}{2} g \tau_{i}^{2}  \;\;\; ,
\end{equation}
with $t_{i}$, and $\tau_{i}$ time lags for 
$i$ sets of particles.  The mean value of $\tau_{i}$ is deduced 
by considering the limit $t \rightarrow \infty$, which leads to 
$h(\infty) = <h_{i}(\infty)> = \sum_{i=1}^{N} \rho_{i} h_{i}(\infty) = g<\tau_{i}^{2}>/2$. 
In this way, the mean time lag $<\tau_{i}>$ is interpreted
as the time at which the growth velocity becomes zero and the maximun 
body height is achieved with respect to the earth's surface $h(0)=0$
(recall that $x(t)=x(0)+v(t)*t+\frac{1}{2}a(t)*t^{2}$
from Newton's mechanics for constant mass systems).  On the other hand, 
$t_{i}$ are the time lags at which the single height contributions 
$h_{i}(t)$ of the $N$ objects or set of particles become zero.

Furthermore,
\begin{equation}\label{eq:velocity}
v(t) = \frac{dh(t)}{dt} 
 =  <\frac{dh_{i}(t)}{dt}>  = 
\sum_{i=1}^{N} \rho_{i} \frac{h_{i}(\infty)}{\tau_{i}} \; sech^{2} \left(\frac{t-t_{i}}{\tau_{i}}\right)  \;\;\; ,
\end{equation}
which for $t \rightarrow \infty$, it gives $v(\infty) \rightarrow 0$
as expected (after completion of the physical growth), and
\begin{equation}\label{eq:mass}
m(t) = < m_{i}(t) >  = \sum_{i=1}^{N} \rho_{i} m_{i}(\infty) \; tanh \left(\frac{t-t_{i}}{\tau_{i}}\right)  \;\;\; .
\end{equation}
It follows inmediately that physics theory predicts that the height
and mass functions reach maximum values (plateau) on age due to the
presence of the hyperbolic tangent functions.  The mass per unit
height $\lambda_{i}(\infty) =m_{i}(\infty)/h_{i}(\infty) \ne 0$
corresponds to the slope of $m_{i}(t)$ versus $h_{i}(t)$.
For systems of variable mass, the center of mass velocity $v(t)$
is not the same as that for the $N$ sets of particles in the 
system $v_{i}$.  In turn, the moving particles may not have  
the same velocities for each state $i$.  

\section{Physics estimates of human growth data}

To this point, the amount of growth parameters in the present physics 
based biometric formulae seems to be large.  However, these 
are not only comparable with those in the alternative asymptotic 
mathematical models tested for the modeling of human anthropometric 
data (see, {\it e.g.}, \cite{Fle99,Col90,Boc76,Kan90,Per02,Jol92,Kar89,Sho91}),
but their number can still be reduced --at least for small $N$ 
sets of particles.

It is easy to verify by a plot that for $y\ge 0$, the monotonicity of the 
hyperbolic tangent implies the relation \cite{Mali01} 
\begin{equation}\label{eq:tanhxy}
tanh \; y \; \le 1 \le \frac{tanh \;(xy)}{tanh \;x} \;\;\; .
\end{equation}
Therefore, using this inequality it is possible to group together model 
parameters to set lower bounds to all functions in this work.  
In fact, by setting $y = \tau/\tau_{i} \rightarrow 1/\epsilon_{i} \ge 0$ 
(independent of $t$) and $x \rightarrow (t-t_{i})/\tau$, such that 
$\tau_{i} \equiv \epsilon_{i}\tau$, one obtains
\begin{equation}\label{eq:tanhxy1}
tanh \;(xy) =
tanh \;\left( \left[\frac{t-t_{i}}{\tau}\right] \; \left[\frac{\tau}{\tau_{i}}\right]\right) \geq
tanh \; \left(\frac{\tau}{\tau_{i}}\right)  
           tanh \; \left(\frac{t-t_{i}}{\tau}\right) 
   = tanh \; y \; tanh \;x \;\;\; .
\end{equation}
Hence Eqs.(\ref{eq:height}) and (\ref{eq:mass}) can be approximated as
\begin{equation}\label{eq:height2}
h(t) \geq \sum_{i=1}^{N} A_{i}\; tanh \left(\frac{t-t_{i}}{\tau}\right) \;\;\; .
\end{equation}
and
\begin{equation}\label{eq:mass1}
m(t) \geq \sum_{i=1}^{N} B_{i} \; tanh \left(\frac{t-t_{i}}{\tau}\right) \;\;\; ,
\end{equation}
where, $\forall i$, it is defined
\begin{equation}\label{eq:ab}
A_{i} \equiv \rho_{i} h_{i}(\infty)\; tanh \left(\frac{1}{\epsilon_{i}}\right)
= \rho_{i} h_{i}(\infty)\; tanh \left(\frac{\tau}{\tau_{i}}\right) 
\;\;\; , \;\;\; 
B_{i} \equiv \lambda_{i}(\infty) A_{i} \;\;\; , 
\end{equation}
with $\lambda$ is mass per unit height as before.
Since the $\tau_{i}$ parameters are related to $h_{i}$ via the Newtown 
equation in Eq.(\ref{eq:height1}), then this leads to the positive
dimesionless quantities 
$\epsilon_{i} = (1/\tau) \sqrt{2 h_{i}(\infty)/g}$. Furthermore,
the momentum flux tranferred to the body by the added $m_{i}$ becomes
$\phi_{i} (t) \equiv v_{i}\;(dm_{i}/dt) = (B_{i}/A_{i})\;v_{i}(t)^{2}$,
similarly to nonrigid systems \cite{Sou02}.

Taking the derivative of Eq.(\ref{eq:height2}) it follows that
\begin{equation}\label{eq:velocity1}
v(t) \geq \frac{1}{\tau} \sum_{i=1}^{N} A_{i} \; 
          sech^{2} \left(\frac{t-t_{i}}{\tau}\right) \;\;\; .
\end{equation}
which is consistent with Eq.(\ref{eq:velocity}) in the lower bound since
$(1/\tau_{i}) sech^{2}(xy) \geq (1/\tau) \; tanh\; y \; sech^{2}\;x$.

In this way, the unkown growth parameters $h_{i}(\infty)$, 
$m_{i}(\infty)$ and $\rho_{i}$ can reduce to $A_{i}$ and $B_{i}$,
which will be referred to as {\it "biological parameters"}.
Together with the time lag $\tau$ and the peak positions
of each $tanh$ function, given by $t_{i}$, these parameters allow 
to describe non-linear percentile curves (or alternatively
z-scores \cite{Fle99}) as shown next.

Selected smoothed empirical percentiles of measured mass (weight)- 
and height-for-age from the boys data based on medical literature 
\cite{nchs} are shown in Figs 2 and 3 by dots, respectively.  
The theoretical curves (full lines) are obtained using 
Eqs.(\ref{eq:mass1}) and (\ref{eq:height2}) assuming 
$N=6$ sets of particles with the biological parameters 
$A_{i}/B_{i}$ of Eq.(\ref{eq:ab}) illustrated in Fig.4.  
All parameter values can easily be derived using the fit 
command of the free GNUPlot data plotting software.  
Real (smoothed and unsmoothed) data curves for boys 
and girls increase monotonically with age independent of 
gender and race 
\cite{Fle99,Col90,Boc76,Kan90,Per02,Jol92,Kar89,Sho91}.
In the inserts, unsmoothed $50\%$ percentile empirical 
points for $w(t)$ and $h(t)$ of Chinese girls ages 6-18 
years \cite{Chan65} are shown for comparison to display 
observed characteristic trends.

In order to show that there is not need for guessing some 
mathematical functions a priori, let us fit the smoothed 
curves rather than the raw data. 
A similar behaviour to the reported smoothed curves is obtained
from the Newton dynamics of systems with variable $m$.  Within
measurements error, the variation of estimates of growth charts
is close to all smoothed empirical data in the figures
(usually measured to the nearest $0.1 cm$ \cite{Xxu02}).
Both the height-for-age and measured mass-for-age tend
to a maximum.  According to these results, it can be argued that
for each plotted point, the variable mass and height
(and growth velocity as seen below) of humans can be 
correlated in relation to physics laws over many years.

At the origin, the boundary conditions $h(0) = 0$ and $m(0) = 0$ in
Eqs.(\ref{eq:height2}) and (\ref{eq:mass1}) lead to the relations
between the biological parameters
\begin{equation}\label{eq:bc1}
A_{1} \; tanh \left(\frac{t_{1}}{\tau}\right) \leq
\sum_{i=2}^{N} A_{i} \; tanh \left(\frac{t_{i}}{\tau}\right)
   \;\;\; ; \;\;\;
B_{1} \; tanh \left(\frac{t_{1}}{\tau}\right) \leq
\sum_{i=2}^{N} B_{i} \; tanh \left(\frac{t_{i}}{\tau}\right) \;\;\; ,
\end{equation}
where $t_{1} < 0$ and $t_{2}, \cdots, t_{N} > 0$ which leads to
reduce further the number of parameter values.  In fact, as
$t \rightarrow \infty$, $tanh \; t_{_{i\; > N}} \rightarrow 1$
and $B_{_{i\; > N}} \rightarrow 0$ for finite $N$ sets of particles
(of added matter $m_{_{i\; \le N}}$).

In humans growth is a natural process characterized by periods
of growth spurts during early infancy and again at puberty.
In Figs.5 and 6 illustrated by dots are the growth velocity 
curves and the measured mass (weight)-for-height curves at different 
percentiles, respectively \cite{nchs}.  The full lines in these 
figures correspond to results obtained from Eq.(\ref{eq:velocity1}) 
and Eqs.(\ref{eq:mass1})-(\ref{eq:height2}), using
parameters as before.  As shown in Fig.5, the growth velocity 
oscillates down as a function of time to vanish 
in the adolescence.  The system acceleration $dv/dt$ changes 
sign over the years due to the $tanh$ term in the derivative 
of Eq.(\ref{eq:velocity1}).  On the other hand,
both the real and the mass-for-height model curves of Fig.6 increase 
monotonically.  The $5^{th}$ percentile in this case is considered 
as an underweight threshold \cite{nchs}.  Such weight-for-height 
standards are used in the study of nutrition status and 
mortality risk in a community \cite{Fri90}.  

Further evidence for a relation between physics 
laws and human growth is given by looking at the 
BMI on age plotted in Fig.7 for selected percentiles.  
On equal grounds, the plotted dots are evaluations 
using the measured mass and heigth on age using the smoothed 
charts of Figs.2 and 3 \cite{nchs} and the full line curves from
$m(t)/h(t)^{2} = <m_{i}(t)>/<h_{i}(t)>^{2}$
with the biological parameters 
in Fig.4.  It can be seen that the present estimates of BMI
are reasonably close to the reported smoothed data.  
BMI fluctuations increase with the error of the 
data for $h^{2}$.  This is reflected in BMI
calculations for younger ages at the higher percentiles.
Theoretical BMI curves at the lower percentiles display 
a minima with increasing age from 3 to about 6 years to 
then rise steadily with age as observed in reality.  
Within the physics approach, this minimium is a consequence
of the growth spurts at early infancy depicted in Fig.5.  
Increasing $N$ in the sums of 
Eqs.(\ref{eq:mass1}) and (\ref{eq:height2}) allows a better 
minima fit at higher percentiles of the BMI-for-age curves.

\section{Remarks}

The present study is not based on arbitrary mathematical 
trial functions to fit the human physical growth data.
All general trends for the observed non-linear percentile curves 
displaying the distribution of different aspects of human physical 
growth from childhood to adolescence have been understood and 
described in relation to physics laws that are daily perceived 
by living on the surface of the earth.  

As shown here theoretically (see Appendix A for details on the derivation 
of the general Newton approach in question), the underlying physics of 
Newtonian mechanics is the bases of the derivation of all equations.
The coefficients at each stage of the
derivations are the consequence of dealing with the differential equation
(\ref{eq:force}) applied to a very dense data set based on human growth.
This also include the fundamental function, $tanh(x)$ appearing throughout
the resulting relations for the body height in Eq.(\ref{eq:height})
, mass in Eq.(\ref{eq:mass}) and BMI thereafter.
Therefore $tanh(x)$ here is not made arbitrarily
to fit a wide range of data from a set of adjustable, weighted coefficients.
On the other hand, the peaked growth velocity of Eq.(\ref{eq:velocity})
depends on the different hyperbolic function $sech(x)$.

Alternative models to estimate growth curves and their derivates, such as the
growth velocity, are based on the use of a non-parametric monotone smoothing
process of the form \cite{Ram02}
\begin{equation}\label{eq:fda}
\mu (t) = \beta_{0}  + \beta_{1} \int_{0}^{t} exp\{W(u)\}du \;\;\; .
\end{equation}
The regression coefficients $\beta_{0,1}$ set the origin and range 
required to fit $\mu$ data including the velocity curves $v(t)$ on age 
to account for the pubertal growth spurt.
The $\mu (t) \rightarrow v(t)$ function is constructed using the
functional data object 
$W(u)$ which does not approximate the measured data directly, but rather
after some exponential transformations (to make $\mu$ positive) and
integration up to $t$ (to make $\mu$ increasing or monotonic).
A relationship between the present physics based approach and the
monotone smoothing can be established as shown explicitly 
in Appendix B.

The tendency to predict and describe growth by normal
curves often obscures individual growth spurt phases that are consistently
present but asynchronous across the population as a whole \cite{Ram02}.
To this end, one can deduce that since
the present growth theory is based on the law of gravity which influences 
individuals, then either a single contribution, or the superposition of 
contributions, on the growth curves are driven by the same general
patterns described as derived by Newton's law in Eq.(\ref{eq:force}).

When dealing with an extended system such as the human body, 
Newton's second law states that the net external force applied 
to the center of mass equals the center of mass acceleration times the mass, namely
${\bf F}_{\it ext} = M {\bf a}_{\it (cm)}$.  
According to Eq.(\ref{eq:force1}) one can then write
$\hat{F}_{\it ext} = <m_{i}v_{i}\; (dv_{i}/dh_{i}) - 2v_{i}^{2} \; (dm_{i}/dh_{i})>  = 
< m_{i}(\infty) a_{i \it (cm)} >$.
Using the average definition in Eq.(\ref{eq:average}), {\it e.g.},$m = < m_{i} >$, 
in conjuction with the definiton of mass-per-unit height
$\lambda_{i}(\infty) \equiv m_{i}(\infty)/h_{i}(\infty)$
and Eqs.(\ref{eq:height2})-(\ref{eq:ab}) it follows that
$h_{i}(t) = (h_{i}(\infty)/m_{i}(\infty))m_{i}(t)$ and
$v_{i}(t) = h_{i}(\infty) (dm_{i}(t)/dt)$.  Hence
$h_{i}(\infty) a_{i \it (cm)} \equiv h_{i}\; (dv_{i}/dt) - 2\; v_{i}^{2}$.
The latter corresponds to the difference between the center of mass 
acceleration of a uniform rod of height $L$ lifted with a variable 
velocity $v$, namely 
$L \times a_{\it (cm)}^{\it rod} = v^{2} + x\; (dv/dt)$, and 
that of a uniform rod of height $L' = L/3$ lifted with constant velocity 
$v' = v$, namely $L' \times a_{\it (cm)}^{\it rod'} = v'^{2}$. 
This superposition may reflect the fact that the center of mass 
of the human body --whose mass distribution can also change by pelvic
rotation and displacement, knee flexion, foot mechanisms-- has not 
a fixed location on time \cite{Ore04}.
There are alternative models based on Newton mechanics to study
mass redistribution throughout the body as a function of age 
and applied forces (see, {\it i.e.}, \cite{Jen89}).
By applying the equation of motion for the human body with variable mass
as discussed here, it could be helpful to understand the extent to
which empirical models by inverse dynamics to derive the torque 
(moment) correspond to prediction as well as their affect on human 
movement.

The primarily interest in human physical growth and its relationship 
to physics laws is to enhance the fit quality to observed data
and to understand the underlying phenomena. 
There is overwhelming biological evidence that continuous
growth processes on earth proceed by multiplicative increments, 
therefore the body size of plants, animals and humans follows a 
lognormal rather a normal distribution (with additive increments) 
\cite{Hux32}.  Thus one should not expect human height distributions 
to be normal but lognormal \cite{Sch02}.  The claim here is not 
modified by either class of $\rho_{i}$ distributions --relating the
biological parameters in Eq.(\ref{eq:ab}).  In principle,
one may associate these parameters to changes in kinetic energy 
of the center of mass of the system \cite{Sou02}, and the thermal 
internal energy, via the integral of $\hat{F}_{\it ext}(t)$ 
in Eq.(\ref{eq:force1}) with respect to 
$x_{\it (cm)}^{rod} = x^{2}/2L - x'^{2}/2L'$.  
 
The relationship between natural laws and human physical growth 
as introduced in this work may be also suitable to understand the 
biometrics of other living species growing vertically such as
Primates \cite{Lei01} (and perhaps leaving trees and plants too).  
However, at this point, it is also worth to ask if the same physics laws
discussed here can also be applied in the case of horizontally growing
terrestial organisms, {\it i.e.}, those organisms growing in an 
essentially weightless environment (aquatic organisms as white-side
Dolphins \cite{Fer96}).  

So the question is: {\it Can horizontal growth be also described 
within the present physics approach in the "absence" of gravity 
($g \rightarrow 0$) as contrary to vertical growth where gravity
influences directly?}

Let us attempt to reply this crucial question and see how the present 
model may provide a unique description for both cases.
Consider the general Eq.(\ref{eq:force1}) for $\hat{F}_{\it ext}$ again 
and set the force of gravity equal to $\hat{F}_{g}=0$. Alternatively, from
Eq.(\ref{eq:force2}) this implies $h_{i}(t) \rightarrow h_{i}(\infty) = const.$ 
which means to compensate out the weight $w(t)$ with the volume dependent 
forces.  Therefore
\begin{equation}\label{eq:forceHor}
\sum_{i=1}^{N} \rho_{i} \; m_{i}(t)\; \frac{dv_{i}(t)}{dt} -
           2 \sum_{i=1}^{N} \rho_{i} \;
     v_{i}(t) \; \frac{dm_{i}(t)}{dt}  = 0 \;\;\; .
\end{equation}
One solution satisfies
\begin{equation}\label{eq:Hor1}
 m_{i}(t)\; \frac{dv_{i}(t)}{dt} = \; 2 v_{i}(t) \; \frac{dm_{i}(t)}{dt}  \;\;\; ,
\end{equation}
or
\begin{equation}\label{eq:Hor2}
 \frac{dv_{i}(t)}{v_{i}(t)} = 2 \; \frac{dm_{i}(t)}{m_{i}(t)} \;\;\; ,
\end{equation}
which, by integration, implies the relation
\begin{equation}\label{eq:Hor3}
 v_{i}(t) \equiv \frac{dx_{i}(t)}{dt} = \gamma m_{i}^{2}(t) \;\;\; ,
\end{equation}
or
\begin{equation}\label{eq:Hor3a}
v(t) = < v_{i}(t) > = \gamma < m_{i}^{2}(t) > \;\;\; ,
\end{equation}
with $\gamma$ a constant for a given aquatic specie and 
for the long $x$-axis of a horizontal structure.

On the other hand, let us consider (and expand in Taylor 
series) the well-known Laird-Gompertz empirical formula for  
fish growth \cite{Fer96,Reg80}.  That is
\begin{eqnarray}\label{eq:Hor4}
\frac{x(t)}{x_{0}} & = & e^{a\; (1-e^{-\alpha t})} \nonumber \\
   & \approx  &   1 + a\; (1-e^{-\alpha t}) \nonumber \\
   & =  &   1 + a\; \left(1- \frac{1}{e^{\alpha t}} \right) \nonumber \\
   & \approx &   1 + a\; \left(1- \frac{1}{1 + \alpha t} \right) \;\;\; 
\end{eqnarray}
where $x(t)$ is the lenght at age $t$ (at zero gravity) and $(a,\alpha)$ 
are specific rates of the exponential growth.  
It follows that the growth velocity $v(t) \equiv dx(t)/dt$ satisfies
\begin{equation}\label{eq:Hor5}
v(t) = \frac{ax_{0}\alpha}{(1+\alpha t)^{2}} =
      \frac{\alpha}{ax_{0}}[(1+a)x_{0} - x(t)]^{2} \;\;\; . 
\end{equation}
Using the Chebyshev's sum inequality $\sum_{k=1}^{n}a_{k}b_{k} \ge 
(1/n)\sum_{k=1}^{n}a_{k}\sum_{k=1}^{n}b_{k}$, such that
$a_{k}=b_{k} = \rho_{i}m_{i} \equiv \rho m_{i}$, then
comparison of Eqs.(\ref{eq:Hor3}) and (\ref{eq:Hor5}) results in
$< m_{i}(t)>^{2} \; \le N\rho \sum_{i=1}^{N}\rho m_{i}(t)^{2} = N\rho <m_{i}^{2}(t)>$ and
\begin{equation}\label{eq:Hor6}
m(t) \le  \left(\frac{N\rho\alpha}{ax_{0}\gamma}\right)^{1/2} |(1+a)x_{0} - x(t)| \;\;\; .
\end{equation}
which for values $x(t) > (1+a)x_{0}$, it implies an approximated linear 
relation between mass and length for horizontal growth (the mass-for-height 
for vertical growth under the influence of gravity is plotted in Fig.6
for comparison).

To this end, collected data reported in \cite{Fer96} for white-side Dolphins 
in the central north pacific ocean suggest that the postnatal dentine thinkness
of Dolphins (used as an index of age in these animals) increases linearly
with the total body length.  It is reasonable to then justify the
linear proportion between the aquatic species mass and length of the type
in Eq.(\ref{eq:Hor6}), as a consequence of the application of physical laws
({\it c.f.}, Eq.({\ref{eq:force1})), in conjuction with the Laird-Gompertz 
growth function ({\it c.f.}, Eq.({\ref{eq:Hor4})).  Relations (\ref{eq:Hor4}) 
and (\ref{eq:Hor6}) also allows to study the BMI for aquatic organisms and
Eq.(\ref{eq:Hor5}) to predict their growth velocity curves in terms of the
square of their mass ({\it c.f.}, Eq.({\ref{eq:Hor3a})).

Therefore, the absence of gravitational forces during growth in 
secondarily aquatic tetrapods (SAT) --backboned, limbed, air-breathing animals
with ancestors that evolved in a terrestrial environment where gravity was a factor
in growth and motion ({\it e.g.}, Dolphins)-- leads to mass and body length trajectories
that, in principle, could be replicated by the present general physics model.
Further multidisciplinary research along these lines is needed
to shed new light on the fundamentals of growth in all classes of
terrestrial organisms using one model based on Newton's second law 
as discussed here. 

\section*{Acknowledgements}

Sincere thanks are due to the Referees for helpful observations.

\section*{APPENDIX A}

Let us follow Refs. \cite{Hal78,Sou02} to demonstrate Newton's law 
in Eq.(\ref{eq:force}) for open systems where the mass varies with 
time and is not constant.  

The net external force acting on the system can be approximated as
\begin{equation}\label{eq:app1}
{\bf F}_{\it ext} =  \frac{d{\bf P}}{dt} \approx 
        \frac{\Delta {\bf P}}{\Delta t}  =
        \frac{{\bf P}_{f} - {\bf P}_{i}}{\Delta t} \;\;\; ,
\end{equation}
where ${\bf P}_{f,i}$ denotes the final and initial system momentum, 
respectively.  As an illustrative example, let us consider a mass $m$ 
moving with velocity ${\bf v}$ that ejects a mass $\Delta m$ during a 
finite time interval $\Delta t$.  One can then write
\begin{equation}\label{eq:app2}
{\bf P}_{i}  =  m {\bf v} \;\;\; ; \;\;\;
{\bf P}_{f}  =  (m - \Delta m)({\bf v} + \Delta {\bf v}) + 
            \Delta m {\bf u}\;\;\; ,
\end{equation}
where ${\bf u}$ is the velocity of the center of mass of the ejected 
mass (not longer in the original system).  The system mass is reduced to 
$m - \Delta m$ and the velocity of the center of the system is changed 
to ${\bf v} + \Delta {\bf v}$.  Therefore, replacing Eq.(\ref{eq:app2})
into (\ref{eq:app1}) leads to
\begin{equation}\label{eq:app3}
{\bf F}_{\it ext} =  m \; \frac{\Delta {\bf v}}{\Delta t} + 
     [ {\bf u} - ({\bf v} + \Delta {\bf v}) ] \;
        \frac{\Delta {\bf m}}{\Delta t} \;\;\; .
\end{equation}
The quantity ${\bf u} - ({\bf v} + \Delta {\bf v})$ is just the 
relative velocity ${\bf v}_{\it rel}$ of the ejected mass with respect 
to the main body.

The change in mass with time, is intrinsically negative in this 
example.  As $\Delta t \rightarrow 0$, then the positive quantity 
$\Delta m/\Delta t$ can be replaced by $- dm/dt$.  Hence,
\begin{equation}\label{eq:app4}
m\; \frac{d{\bf v}}{dt} = {\bf F}_{\it ext}  +  
    {\bf v}_{\it rel} \; \frac{dm}{dt} \;\;\; .
\end{equation}
The term ${\bf v}_{\it rel}(dm/dt)$ is the rate at which momentum is
being transferred into (or out of) the system by the mass that the system
has ejected (or gained).  It is interpreted as the force exerted on the
system by the mass that leaves (or joins) the system.

When the volume is allowed to vary, a term accounting
for the additional momentum that the system acquires as a result
of its changing volume (distict from the momentum flux
$\phi (t) \equiv {\bf v}_{\it rel}\;(dm/dt)$) needs to be added 
to Eq.(\ref{eq:app4}) \cite{Tie69}.  In the model for BMI this is given 
by the volume term in Eq.(\ref{eq:force2}).

\section*{APPENDIX B}

The computation of $W(u)$ is difficult and not unique. Its basis expansion
needs to be estimated iteratively.  Some tips on monotone smoothing as
given in \cite{Ram02} suggest to use $n$-order B-spline basis functions
for $W(u)$.  Within this functional data analysis (FDA) it is
necessary to know a-priori the right
behaviour of $\mu(t)$ in order to calibrate the smoothing of growth data
with the smallest possible number of knots.

A relationship between the present physics based approach and the
monotone smoothing using FDA is given next.
Let us expand Eq.(\ref{eq:velocity1}) for the growth velocity
in Taylor series for $sech(x)$, such that
\begin{eqnarray}\label{eq:taylor}
v(t) & \geq & \frac{1}{\tau} \sum_{i=1}^{N} A_{i} \; 
 \left(1 - \frac{\chi_{i}^{2}(t)}{2} + \frac{5\chi_{i}^{4}(t)}{24} - \cdots\right) 
 \left(1 - \frac{\chi_{i}^{2}(t)}{2} + \frac{5\chi_{i}^{4}(t)}{24} - \cdots\right) 
     \;\;\; , \nonumber \\
   & \geq & \frac{1}{\tau} \sum_{i=1}^{N} A_{i} \;
 \left(1 - \frac{\chi_{i}^{2}(t)}{2} - \frac{\chi_{i}^{2}(t)}{2} + \Theta [\chi_{i}^{4}(t)]\right) 
     \;\;\; , \nonumber \\
   & \approx & \frac{1}{\tau} \sum_{i=1}^{N} A_{i} \;
 \left(1 - \chi_{i}^{2}(t)\right)  \;\;\; ; \;\;\; (|\chi_{i}| < \pi/2)
     \;\;\; , \nonumber \\
   & \approx & \frac{1}{\tau} \sum_{i=1}^{N} A_{i} \; exp\{-\chi(t)^{2}\} \;\;\; ; \;\;\;
    (|\chi_{i}| < \pi/2 < \infty) \;\;\; . 
\end{eqnarray}
In the above 
\begin{equation}\label{eq:xt}
\chi(t) \equiv \left(\frac{t-t_{i}}{\tau}\right) \;\;\; ,
\end{equation}
and $4th$-order terms in the expansion of $sech (x)$ and 
$exp (x) \approx 1-x^{2}$ (or alternatively terms of the $6th$-order since, 
to a good approximation, $(2/3)x^{4} \approx x^{4}/2!$ are neglected.

In the continuous limit ({\it i.e.}, $\sum_{i} \rightarrow \int du$), the 
discretized growth velocity function in Eq.(\ref{eq:taylor}) --which varies
in terms of the $i$ index from $1$ to $N$, it becomes 
\begin{equation}\label{eq:int}
v(t) \approx 
\frac{A}{\tau} \int^{(t-t_{N})/\tau}_{(t-t_{1})/\tau} 
       exp\{-u^{2}\}\; du = \frac{A}{\tau} 
  \left\{  \int^{0}_{t/\tau} exp\{-u^{2}\}\; du  +
    \int^{(t-t_{N})/\tau}_{0} exp\{-u^{2}\}\; du  \right\} \;\;\; .
\end{equation}
The integral on the right of the sum corresponds to the well-known error 
function $erf[(t-t_{N})/\tau]$ for finite $t < t_{N}$ and large $\tau$,
with $t_{N}$ the expected value and $\tau$ the standard deviation.
For simplicity an uniform growth due to the $N$ added mass
objects and approximated $t_{1} \rightarrow 0$ and $A_{i}\rightarrow A$ 
is assumed.

Therefore a comparison of Eqs.(\ref{eq:fda}) and (\ref{eq:int}) 
imply these associations 
\begin{eqnarray}\label{eq:assoc}
\beta_{0} & \rightarrow & \left(\frac{A}{\tau}\right) erf[\frac{-t_{N}}{\tau}]  \;\;\; , \nonumber \\
\beta_{1} & \rightarrow & - \left(\frac{A}{\tau}\right)  \;\;\; , \nonumber \\
W(u) & \rightarrow & - u^{2}  \;\;\; .
\end{eqnarray}
Therefore, a link to FDA and the present approach such that $v(t) > 0$
follows by setting $A< 0$ and
$-u^{2} = W(u) = \sum_{i=0}^{n} N_{i,p}(u)P_{i}$ where the sum gives the
B-spline function of degree p with $P_{i}$ the contact points and $u$ the
knot vector.  The function $W(u)$ in Eq.(\ref{eq:int})
leads to the continuous normal (or Gauss) distribution with zero mean
and variance of one.

\newpage

\begin{figure}[!t]
\vspace{4.0cm}
\includegraphics[width=13.5cm]{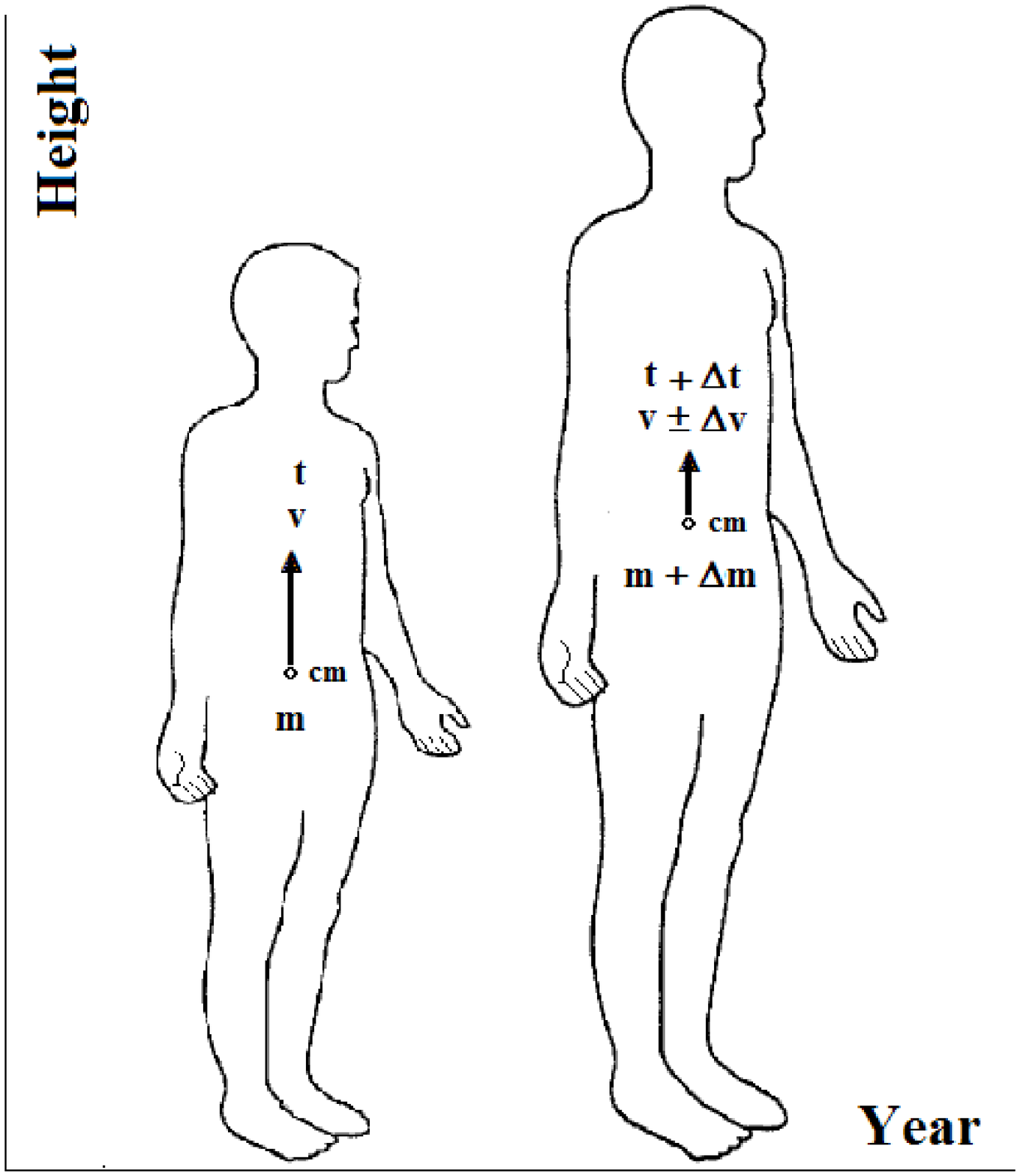}%
\caption{
Representation of human physical growth as a dynamical physics system 
of variable mass (and volume).
}
\end{figure}

\begin{figure}[!t]
\vspace{1.0cm}
\includegraphics[width=10.8cm]{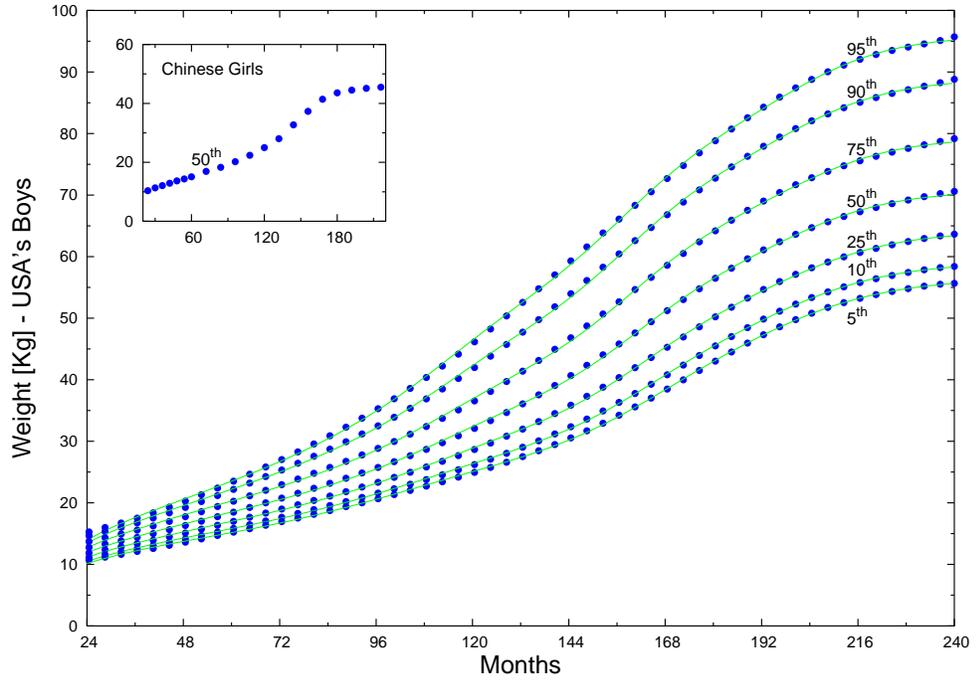}%
\caption{
Measured mass (weight)-for-age for boys at selected percentiles.
Real data (dots) using smoothed charts in \cite{nchs} 
and theoretical results (full lines) from Eq.(\ref{eq:mass1}). 
Unsmoothed $50\%$ percentile empirical points for girls (of 
a different nationality) \cite{Chan65} are shown
in the insert for comparison. 
}
\end{figure}

\begin{figure}[!t]
\includegraphics[width=10.8cm]{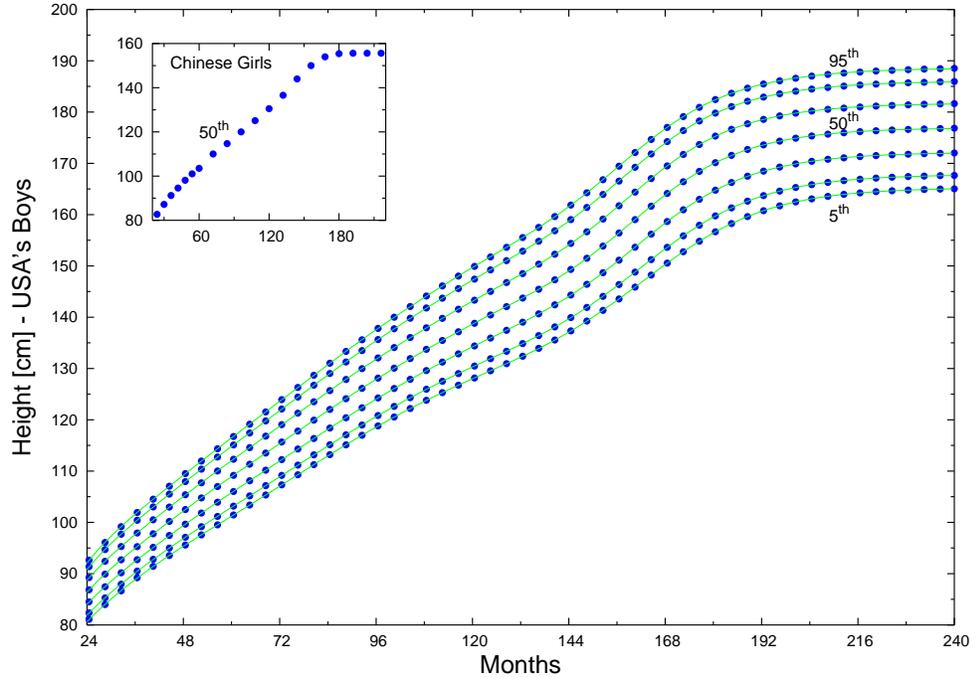}%
\caption{
Height-for-age at percentiles as in Fig.2.  
Real data (dots) using smoothed charts in \cite{nchs} 
and theoretical results (full lines) from Eq.(\ref{eq:height2}). 
}
\end{figure}

\begin{figure}[!t]
\includegraphics[width=13.0cm]{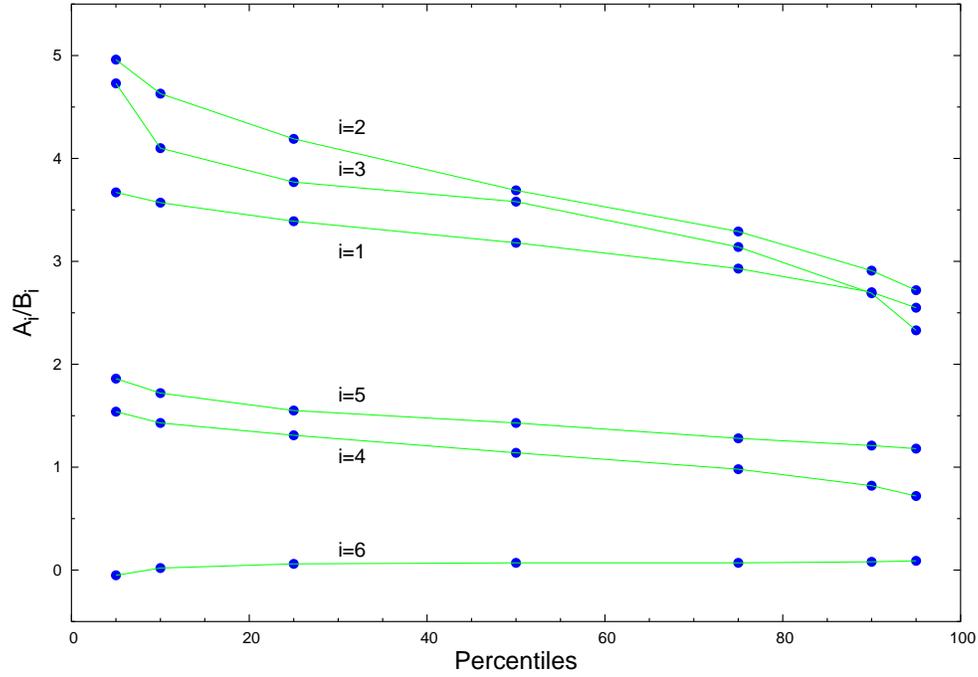}%
\caption{
Inverse of the mass per unit height parameter 
$\lambda_{i}$ ($i=1, \cdots, N$) of Eq.(\ref{eq:ab})
used throughout the physics based calculations.
Full lines are used to guide the eyes.
}
\end{figure}

\begin{figure}[!t]
\includegraphics[width=13.0cm]{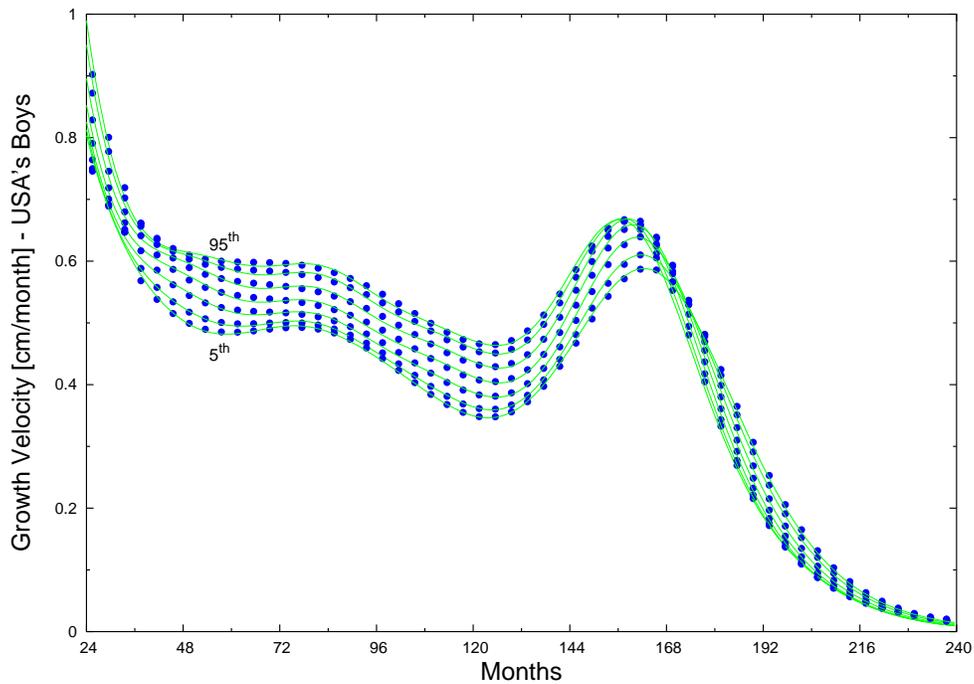}%
\caption{
Growth velocity curves on age at percentiles as in Fig.2.  
Real data (dots) using smoothed charts in \cite{nchs} 
and theoretical results (full lines) from Eq.(\ref{eq:velocity1}).
}
\end{figure}

\begin{figure}[!t]
\includegraphics[width=13.0cm]{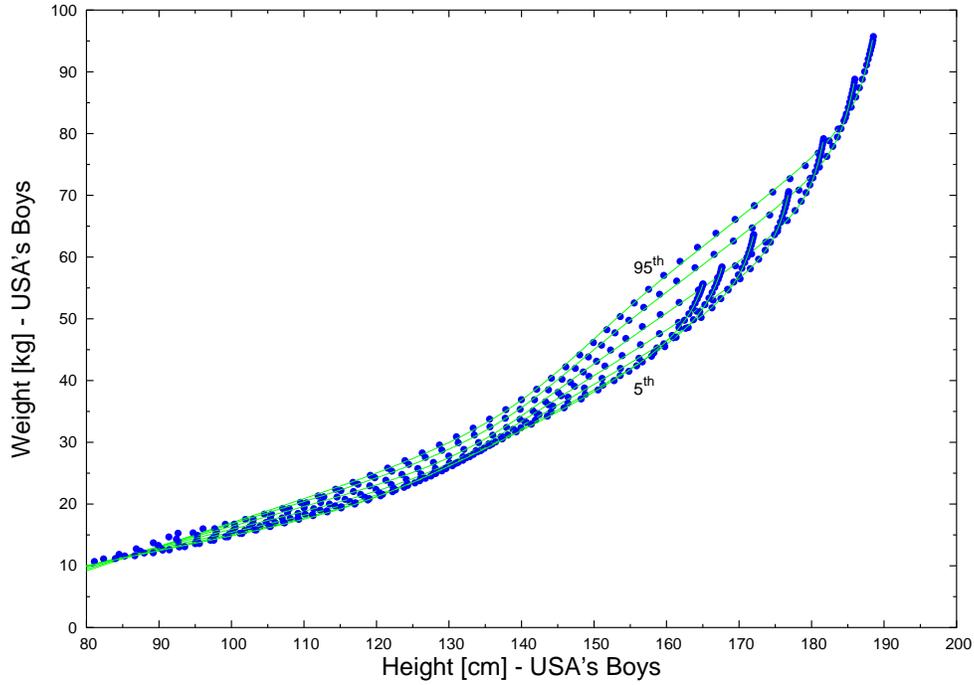}%
\caption{
Measured mass (weight)-for-height at percentiles as in Fig.2.  
Real data (dots) using smoothed charts in \cite{nchs} and theoretical 
results (full lines) from Eqs.(\ref{eq:mass1})-(\ref{eq:height2}).
}
\end{figure}

\begin{figure}[!t]
\includegraphics[width=13.0cm]{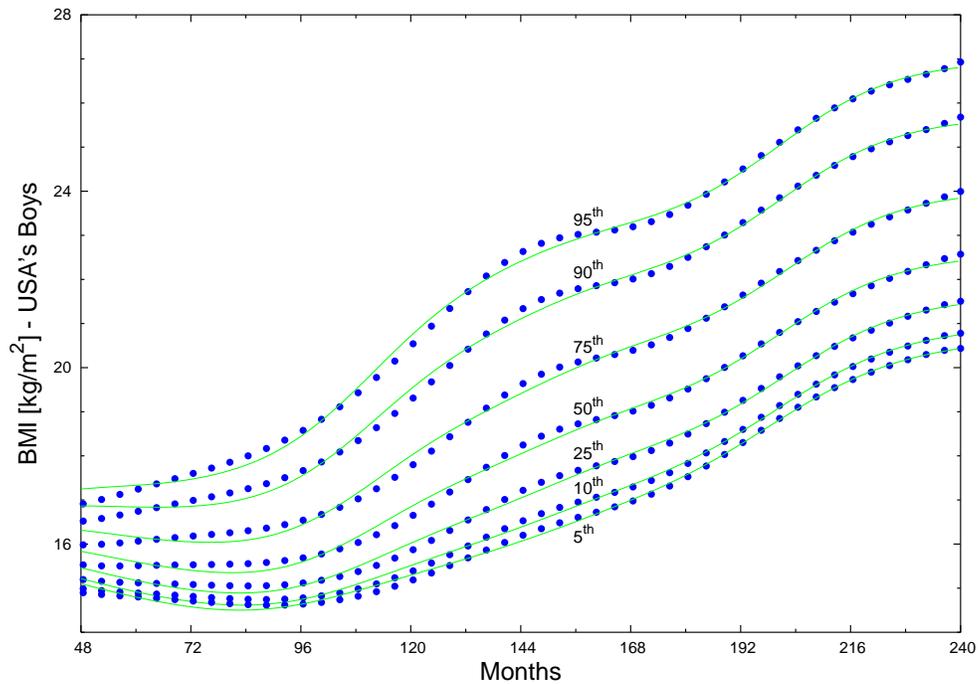}%
\caption{
Body Mass Index at percentiles as in Fig.2.  
Real data (dots) using smoothed charts in \cite{nchs} and theoretical 
results (full lines).
}
\end{figure}
 
\end{document}